\documentclass[aps,prl,reprint,groupedaddress]{revtex4-1}

	\usepackage{amsmath}%,amssymb} 
	\usepackage{makeidx}
	\usepackage{amsfonts}
	\usepackage[ansinew]{inputenc}
	\usepackage[usenames,dvipsnames]{pstricks}
	\usepackage{subfigure}
	\usepackage{epsfig}
	\usepackage{pst-grad} % For gradients
	\usepackage{pst-plot} % For axes
	\usepackage[colorlinks,hyperindex]{hyperref}
	\hypersetup
	{
		colorlinks,%
		citecolor=black,%
		linkcolor=black,%
		urlcolor=black,%
	}
\makeindex

\begin{document}

\title{Competing interactions in artificial spin chains}

\author{V.-D. Nguyen$^{1,2}$, Y. Perrin$^{1,2}$, S. Le Denmat$^{1,2}$, B. Canals$^{1,2}$, N. Rougemaille$^{1,2}$}

\affiliation{
$^1$ CNRS, Inst NEEL, F-38000 Grenoble, France\\
$^2$ Univ. Grenoble Alpes, Inst NEEL, F-38000 Grenoble, France\\
}

\date{\today}

\begin{abstract}

The low-energy magnetic configurations of artificial frustrated spin chains are investigated using magnetic force microscopy and micromagnetic simulations. Contrary to most studies on two-dimensional artificial spin systems where frustration arises from the lattice geometry, here magnetic frustration originates from competing interactions between neighboring spins. By tuning continuously the strength and sign of these interactions, we show that different magnetic phases can be stabilized. Comparison between our experimental findings and predictions from the one-dimensional Anisotropic Next-Nearest-Neighbor Ising (ANNNI) model reveals that artificial frustrated spin chains have a richer phase diagram than initially expected. Besides the observation of several magnetic orders and the potential extension of this work to highly-degenerated artificial spin chains, our results suggest that the micromagnetic nature of the individual magnetic elements allows observation of metastable spin configurations.

\end{abstract}


\maketitle

\bigskip

Artificial spin ice systems have been introduced about a decade ago \cite{Tanaka2006, Wang2006} as a powerful mean to explore frustrated magnetism experimentally, in a controlled manner \cite{Nisoli2013}.
First designed to investigate the rich physics of spin ice materials \cite{Harris1997}, they offer the advantage of being tunable at will. Besides their tunability, one of the main interests of fabricating artificial spin ice systems is the capability to spatially resolve their spin configurations using magnetic imaging techniques. This allows to visualize in real space collective magnetic phenomena often associated to highly frustrated magnets \cite{Book2009}. 
For instance, artificial spin ice systems permit the evidence of spin liquid states \cite{Qi2008, Rougemaille2011, Anghinolfi2015, Sendetskyi2016}, Coulomb phases \cite{Perrin2016}, complex magnetic ordering \cite{Zhang2012, Chioar2014, Chioar2016}, charge crystallisation \cite{Rougemaille2011, Zhang2013, Montaigne2014, Chioar2014_2, Drisko2015}, monopole-like excitations \cite{Ladak2010, Mengotti2011, Morgan2011, Phatak2011} and spin fragmentation \cite{Canals2016}.

\begin{figure}[h!]
\centering
\includegraphics[width=7cm]{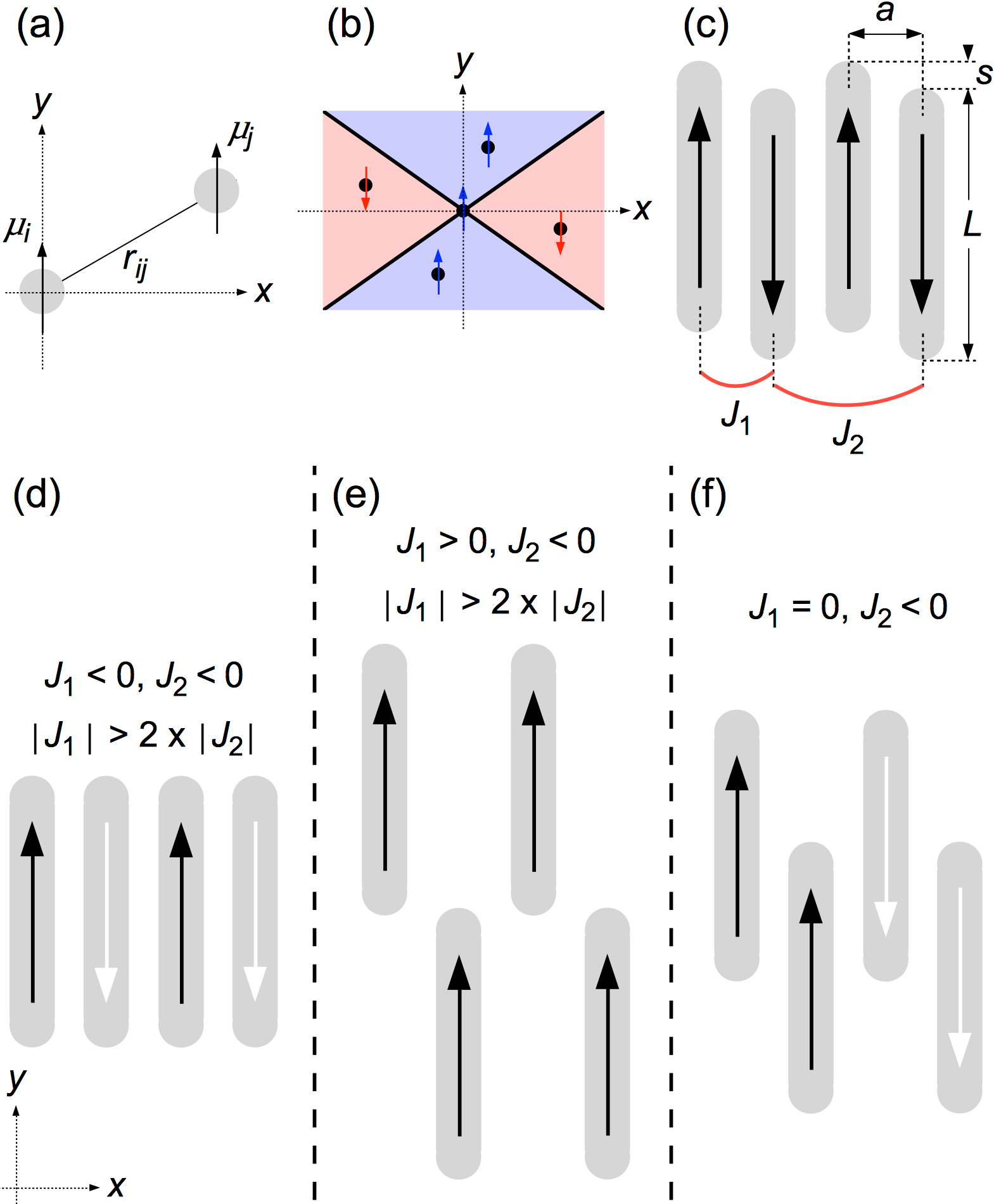}
\caption{(a) Schematics of two magnetic moments $\mu_i$ and $\mu_j$ pointing along the $y$ axis and separated by a distance $r_{ij}$. (b) Diagram illustrating that the dipolar interaction favor ferromagnetic (blue region) or antiferromagnetic (red region) coupling between two magnetic moments depending on their relative position in the $xy$ plane. (c). Schematics showing the different geometrical parameters and coupling strengths relevant in this work. (d-f) Different possible magnetic orders predicted by the one-dimensional ANNNI model depending on the relative strength of the coupling coefficients. $s=0$ (d), $s=L$ (e) and $s<L$ (f). Arrows indicate the magnetization direction.}
\label{fig:principle} 
\end{figure}

Due to their correspondence to natural spin ice materials, most studies so far on artificial spin ice systems have been focused on the two-dimensional square \cite{Wang2006, Perrin2016, Drisko2015, Morgan2011, Phatak2011, Farhan2013, Morgan2013, Kapaklis2012, Porro2013} and kagome \cite{Tanaka2006, Qi2008, Rougemaille2011, Anghinolfi2015, Sendetskyi2016, Zhang2013, Montaigne2014, Chioar2014_2, Drisko2015, Canals2016, Ladak2010, Mengotti2011} geometries. 
In these artificial spin systems, the magnetic moments are Ising-like variables, lying within the lattice plane and pointing locally along the angle bisectors of the checkerboard and kagome lattices, respectively. 
The interaction between nearest neighbors being ferromagnetic, frustration in the artificial square and kagome spin ices is of geometrical origin: the system is not able to satisfy all its pairwise magnetic interactions simultaneously because of the lattice geometry that propagates conflicting information. 

Here, we explore the low-energy magnetic states of artificial spin chains in which frustration is not induced by the underlying geometry, but instead by competing interactions between neighboring elements.
To do so, we follow the strategy developed for artificial Ising chains, where a series of magnetic islands are coupled through magnetostatics \cite{Cowburn2002, Imre2003, Cowburn2000, Bader2002}, and combine it with the idea proposed for the uniaxial triangular \cite{Ke2008, Zhang2011} and square \cite{Arnalds2016} Ising lattices to impose competing interactions.
More specifically, we use the angular dependence of the dipolar interaction coupling neighboring in-plane magnetized nanomagnets to tune the sign and strength of this interaction. 

To illustrate the influence of the angular dependence of the dipolar interaction, we first consider two (classical) magnetic moments $\vec \mu_i$ and $\vec \mu_j$  having an Ising-type degree of freedom, separated by a distance $r_{ij}$ [see Fig.~\ref{fig:principle}(a), where the Ising spins point along the $y$ direction]. 
Depending on the angle between the $y$ direction and the $\vec r_{ij}$ vector, the dipolar interaction favor either a parallel or an antiparallel alignment of the two magnetic moments, as shown in Figure~\ref{fig:principle}(b).

Based on this simple property, we fabricated artificial spin chains from in-plane magnetized, elongated nanomagnets [see Fig.~\ref{fig:principle}(c)], interacting through magnetostatics. 
The aspect ratio of these nanomagnets is such that shape anisotropy determines the magnetization direction, so that each nanomagnet can be considered as an Ising pseudo-spin. 
Arranging these elongated nanomagnets on a unidimensional chain oriented along the $x$ axis, while shifting periodically half of the nanomagnets along the $y$ direction, can be used to control the coupling strength between nearest ($J_1$) and next-nearest ($J_2$) neighbors [see Fig.~\ref{fig:principle}(d-f)]. 
In particular, the vertical shift $s$ [see Fig.~\ref{fig:principle}(c)] allows a fine tuning of the $J_1$ coupling strength, both in amplitude and sign. 
For instance, the condition $s=0$ gives rise to an antiferromagnetic coupling ($J_1 <0$) between nearest neighbors [see Fig.~\ref{fig:principle}(d)], while imposing $s=L$, where $L$ is the length of the nanomagnets, leads to a ferromagnetic coupling ($J_1 >0$) [see Fig.~\ref{fig:principle}(e)]. 
Consequently, intermediate situations can be reached where $s$ is such that $J_1 =0$ [see Fig.~\ref{fig:principle}(f)], or $|J_1| = 2 \times |J_2|$ for example. 
This simple geometrical parameter $s$ thus permits the investigation of different magnetic scenarios, and in particular cases where interactions  between neighboring nanomagnets compete. Comparison with predictions from the one-dimensional Anisotropic Next-Nearest-Neighbor Ising (ANNNI) model \cite{Redner1980, Selke1988} can then done. 

\begin{figure}[h!]
\centering    
\includegraphics[width=6.5cm]{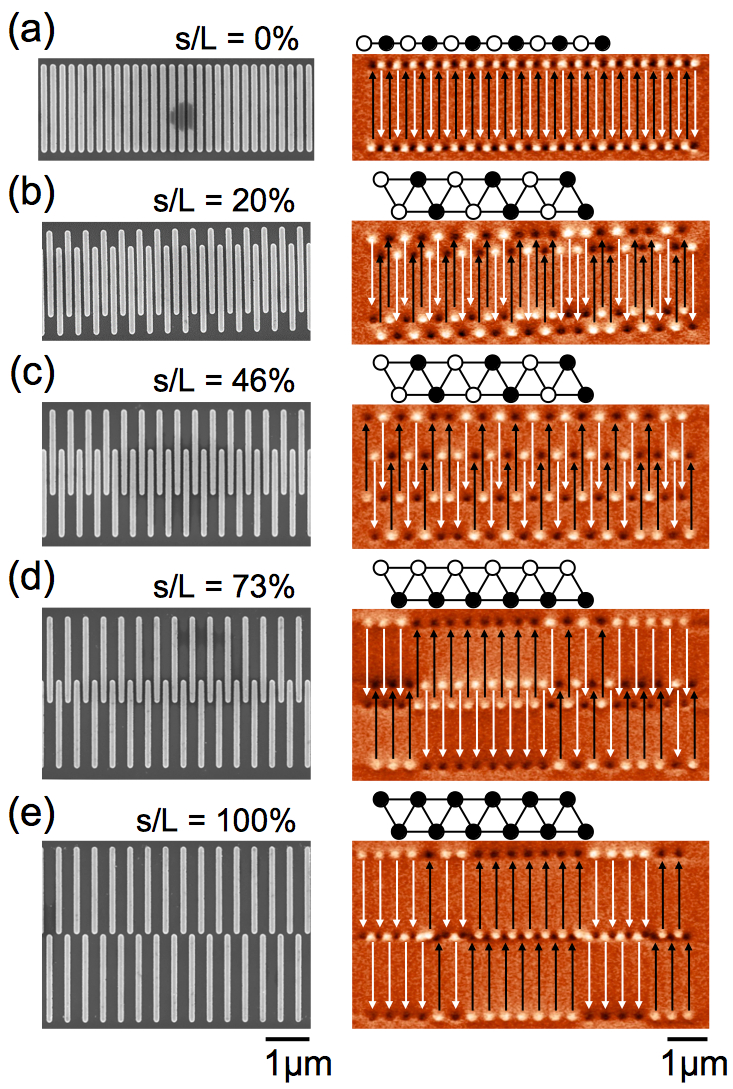}
\caption{SEM (left) and MFM (right) images of demagnetized spin chains for different values of the $s/L$ ratio. On the MFM images, the magnetic contrast appears in black and white, while the black and white arrows indicate the local direction of magnetization deduced from this contrast. The sketches above the MFM images illustrate the expected ground state configuration where the black and white circles code for a spin pointing upwards and downwards, respectively. The three magnetic phases observed experimentally correspond to an antiferromagnetic ordering [(a) and (d)], an antiferromagnetic dimer state [(b) and (c)] and a ferromagnetic phase (e).}
\label{fig:MFM}
\end{figure}

The spin chains were made of permalloy nanomagnets having typical dimensions of $150 \times 2250 \times 30$ nm$^3$, i.e. with an aspect ratio of 15. 
To ensure a significant coupling strength between neighboring nanomagnets, the lattice parameter $a$ [see Fig.~\ref{fig:principle}(c)] was set to 225 nm, leading to a gap between two adjacent elements of only 75 nm. 
Each chain is composed of 40 nanomagnets and the vertical shift $s$ is varied from 0 to 2250 nm by steps of 150 nm. 
The chains have been patterned using e-beam lithography on a Si substrate. 
A 30 nm-thick permalloy film was subsequently deposited using e-beam evaporation and capped with 3 nm of Al to prevent the surface from oxidation. 
The chains were finally obtained using a conventional lift-off process. 
Figure~\ref{fig:MFM} show typical scanning electron microscopy (SEM) images of different chains.

In order to bring our spin chains into a low-energy magnetic configuration, the sample has been demagnetized using a long field protocol. 
Essentially, an in-plane magnetic field is applied and ramped down from 250 mT to 0 in 80 hours using an oscillating, linearly damped current in an electromagnet, while rotating the sample. 
Magnetic force microscopy (MFM) has been then used to image the spin configurations of our demagnetized chains. Figure ~\ref{fig:MFM} shows typical MFM images for spin chains characterized by a shift of $s/L$=0\%, 20\%, 46\%, 73\% and 100\%. 
The black and white contrasts reveal the north and south poles of each nanomagnet. 
Magnetic contrast only appears at the two extremities of the nanomagnets, confirming their single magnetic domain state. 
From this contrast, each pseudo-spin can be unambiguously defined.

We first examine the case of unshifted spin chains ($s/L=0$). 
Within a point dipole approximation, $J_{1}, J_{2} < 0$ and $J_{1}=8\times J_{2}$. 
We thus expect from the one-dimensional ANNNI model to observe an antiferromagnetic ordering of the artificial spin chains after the demagnetization protocol. 
Indeed, our MFM measurements reveal that the unshifted spin chains are in their ground state [Fig.~\ref{fig:MFM}(a)].
Although intuitive, this result is in fact surprising as perfect ordering is found over 40 spins. 
Despite the twofold degeneracy of the ground state, the spin chain was able to eliminate domain walls separating anti-phase domains during the demagnetization protocol. This is in contrast with what was observed in other works where the correlation length was much smaller  \cite{Cowburn2002}. 
Our result is however similar to what was found in artificial spin chains where the ground state degeneracy was intentionally broken \cite{Imre2006} or when the shape of the nanomagnets was made asymmetric \cite{Imre2003}.
Thus, our protocol efficiently brings our artificial spin chains into their low-energy magnetic configurations.

We now study the influence of the vertical shift $s$ on the magnetic configurations observed after demagnetization. 
Results are reported in Fig.~\ref{fig:MFM}(b-e) for the four different ratios $s/L=20\%$, $46\%$, $73\%$ and $100\%$. 
Spin chains made of fully shifted nanomagnets ($s/L=100\%$) show large ferromagnetic domains separated by a few magnetic defects, i.e. magnetic domain walls [Fig.~\ref{fig:MFM}(e)].
In that case, if the first and second neighbor coupling strengths have opposite signs ($J_1>0$ and $J_2<0$), $J_1$ is larger in absolute value and imposes a ferromagnetic order.
However, as we will see below, this result is surprising as it contradicts predictions from the ANNNI model \cite{Selke1988}.

When the shift $s$ is only slightly increased however, an antiferromagnetic dimer phase (i.e. an alternating arrangement of two spins pointing upwards and two spins pointing downwards) is favored to accommodate contradictory information between neighboring elements. 
This is the case for the ratio $s/L=20\%$ [Fig.~\ref{fig:MFM}(b)], where $J_1 \sim J_2 < 0$, consistently with what is expected from the ANNNI model \cite{Selke1988}. 
We note that the same magnetic configuration is obtained for a ratio $s/L$ approaching 50$\%$, where $J_1\sim 0$ and $J_2<0$ [Fig.~\ref{fig:MFM}(c)]. 
This dimer phase would be also obtained for large shifts ($s \gg L$). 
In these two cases, $J_1$ becomes negligible compared to $J_2$, which remains unaffected by the vertical shift $s$. 
In other words, intermediate ($s/L \sim 50\%$) and large ($s \gg L$) shifts also lead to two antiferromagnetic, weakly coupled spin chains, thus forming an antiferromagnetic dimer phase, as expected from the ANNNI model \cite{Selke1988}. 

Surprisingly, between the antiferromagnetic dimer phase and the ferromagnetic order, our spin chains exhibit a transition to an intermediate conventional antiferromagnetic state [Fig.~\ref{fig:MFM}(d)], similar to the one observed for unshifted spin chains [Fig.~\ref{fig:MFM}(a)]. 
The magnetic configuration is puzzling in the sense that the energy of the system is highly increased by the formation of head-to-head and tail-to-tail local configurations. 
As we will see below, we interpret the existence of this intermediate phase (like the ferromagnetic order) as a signature of an out-of-equilibrium physics induced by the demagnetization protocol and the micromagnetic degree of freedom present at the nanomagnets' extremities.

Another way to visualize our experimental findings is to represent the nearest-neighbor spin-spin correlation function for all the spin chains we have fabricated. 
In particular, this quantity allows estimate of how far a given configuration is from the expected ground state and how large is the shift window in which a given magnetic phase is observed experimentally.
The nearest-neighbor spin-spin correlation coefficient is defined as $C_{NN}$=$ \langle \sigma_i \sigma_{i+1} \rangle$, where $\sigma_i$ is an Ising variable ($\pm1$) coding for the spin state of the magnetic moment residing on site $i$ and where $\langle \rangle$ means spatial average over the entire chain. 
Consequently, $C_{NN}$=-1 corresponds to a perfect antiferromagnetic order, while $C_{NN}$=+1 means ferromagnetic order. 
A zero $C_{NN}$ value can be obtained if there is no spin-spin correlations in the system or, more relevant in the present case, if an antiferromagnetic dimer phase is stabilized.
Similarly, the next nearest-neighbor spin-spin correlation coefficient is defined as $C_{NNN}$=$ \langle \sigma_i \sigma_{i+2} \rangle$.

\begin{figure}[h!]
\centering
\includegraphics[width=7cm]{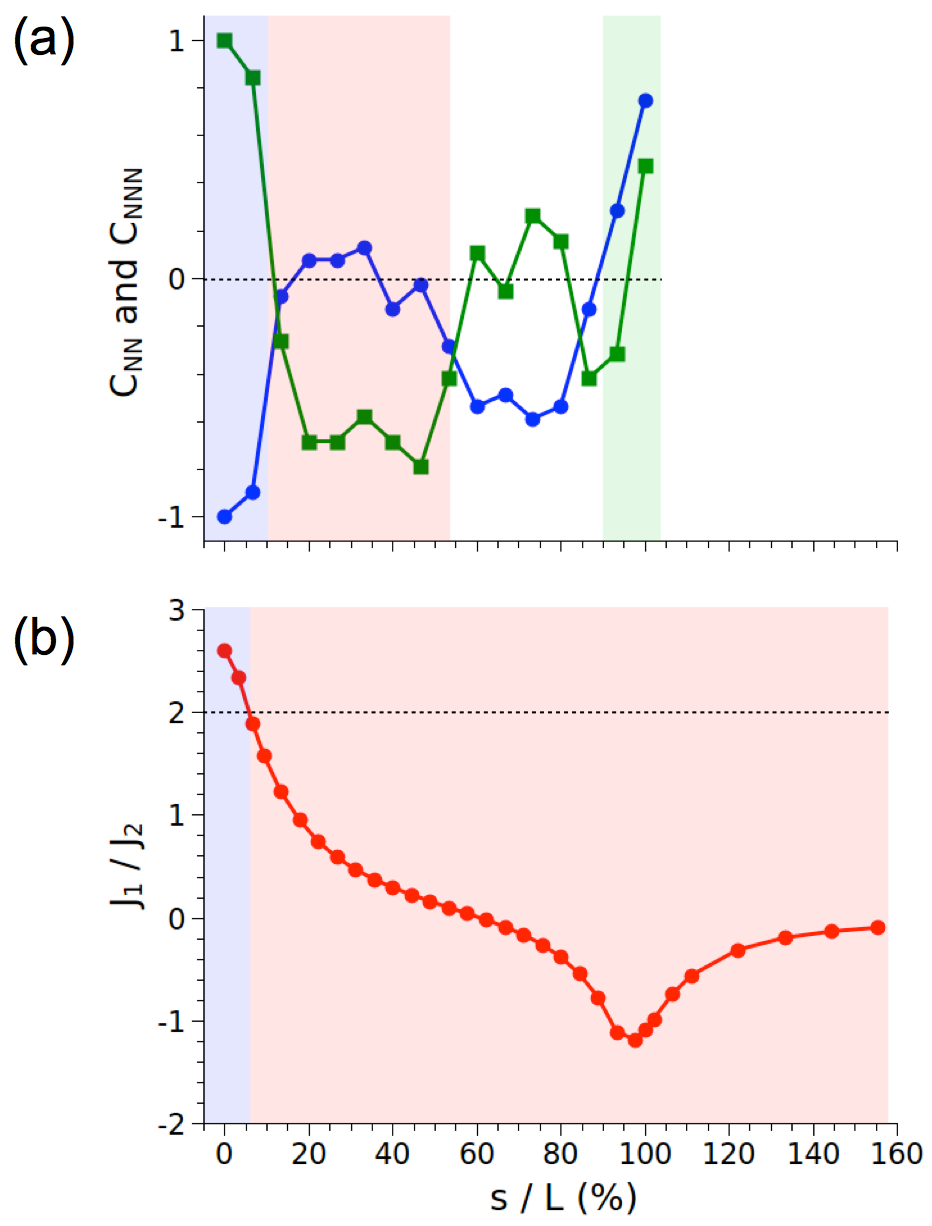} 
\caption{(a) Nearest (blue dots) and next-nearest (green squares) neighbor spin-spin correlations as a function of the $s/L$ ratio deduced from the MFM images reported in Fig.~\ref{fig:MFM}. (b) Ratio of the $J_1$ and $J_2$ coupling strengths as a function of $s/L$ deduced from micromagnetic simulations. Colored regions in both graphes represent the magnetic configurations observed experimentally or expected from the ANNNI model: the antiferromagnetic order (light blue and white), the antiferromagnetic dimer phase (light red) and the ferromagnetic state (light green).} 
\label{fig:phase}
\end{figure}

The $C_{NN}$ and $C_{NNN}$ values deduced from the MFM images obtained after the demagnetization protocol (blue and green curves, respectively) are reported in Fig.~\ref{fig:phase}(a). 
Four different regions can be identified depending on the $s/L$ ratio, corresponding to the four magnetic states described above: a first antiferromagnetic state (light blue region), the antiferromagnetic dimer phase (light red region), a second antiferromagnetic state (in white), and the ferromagnetic order (light green region). 

To interpret our results, we compare the magnetic phases obtained experimentally with those predicted by the ANNNI model.
To do so, we computed the $J_1$, $J_2$ coupling strengths using micromagnetic simulations \cite{OOMMF} by comparing the energy of a pair of nanomagnets in a ferromagnetic and in an antiferromagnetic configuration, for different vertical shifts $s$. 
The mesh size was set to $1\times 1\times 30$ nm$^3$ to limit effects from numerical roughness on the energy estimate. 
Material parameters are those commonly used for permalloy: $\mu_{0}M_S$=1.0053 T, $A$=10 pJ/m, where $M_S$ and $A$ are the spontaneous magnetization and exchange stiffness, respectively.  
Magnetocrystalline anisotropy is set to zero. 
Results of the calculations are reported in Fig.~\ref{fig:phase}(b), where the $J_1 / J_2$ ratio is plotted as a function of $s/L$.

\begin{figure}[h!]
  \centering    
  \includegraphics[width=8cm]{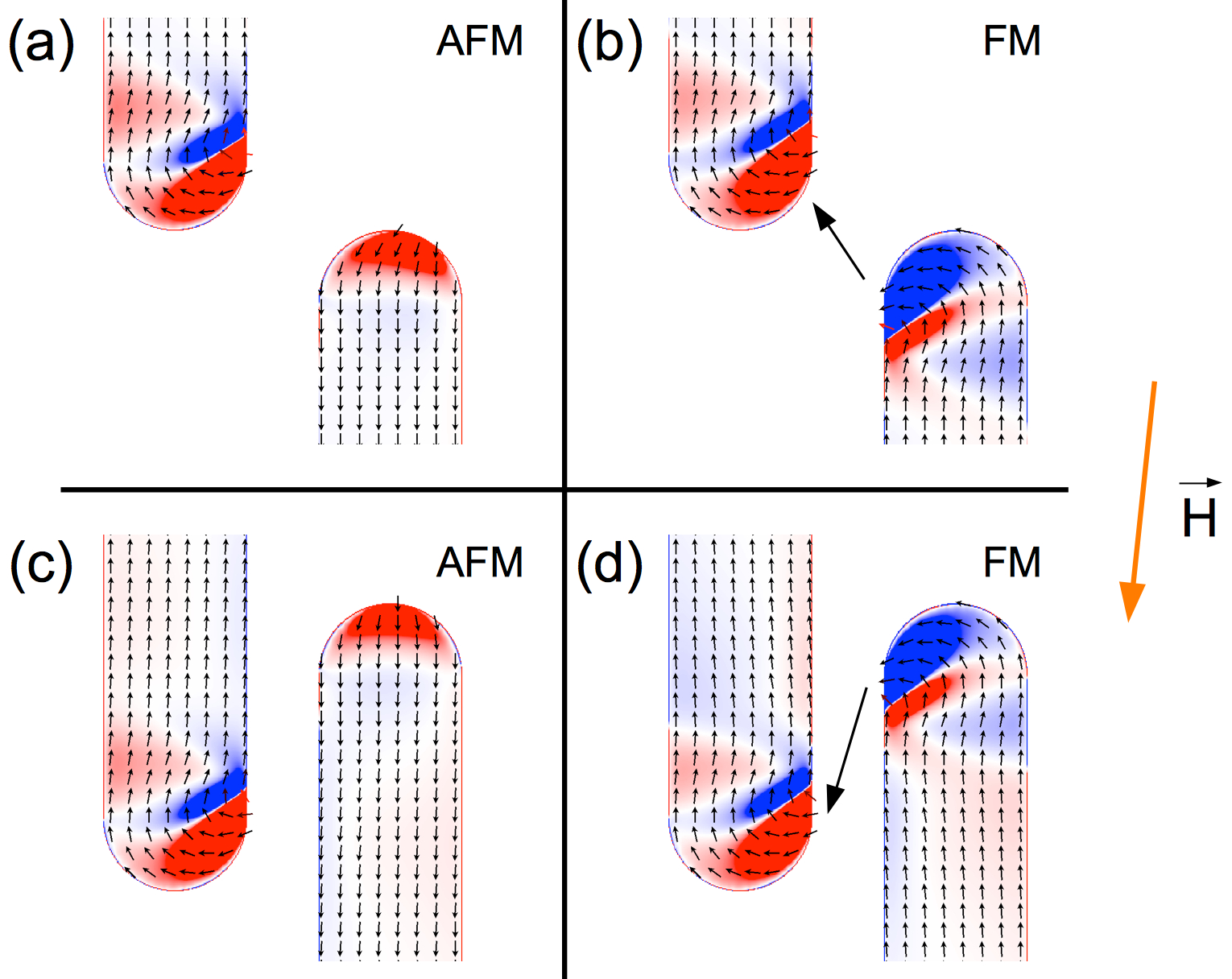}
  \caption{Micromagnetic texture of two neighboring nanomagnets under an applied magnetic field (orange arrow) in four different configurations: for an antiferromagnetic and a ferromagnetic state when $s/L=100\%$ (a-b) and when $s/L=80\%$ (c-d), with $L=2250$ nm. The black arrow in (b) and (d) indicates the dipolar field between the two nanomagnets. The red and blue contrast represents the divergence of the magnetization vector.
}
\label{fig:micromag}
\end{figure}

In the ANNNI model, a magnetic transition is expected when the condition $|J_1| / |J_2|$ = 2 is reached \cite{Selke1988}: an antiferromagnetic configuration for $|J_1| / |J_2| > 2$ and an antiferromagnetic dimer state for $|J_1| / |J_2| < 2$. In Fig.~\ref{fig:phase}(b), we thus identify two different regions.
In light blue, $J_1 / J_2 >2$, meaning that although both couplings favor an antiferromagnetic alignment of the nanomagnets, $J_1$ drives the ordering (antiferromagnetic state). In light red, $|J_1| / |J_2| <2$, the situation is reversed and $J_2$ drives the ordering (antiferromagnetic dimer phase). 
If our measurements are well described by the ANNNI model when $s/L < 50\%$, larger $s/L$ ratios lead experimentally to two magnetic phases that are unexpected.

In the following, we thus investigate the origin of these unexpected antiferromagnetic (white region) and ferromagnetic (green region) phases.
To do so, we performed micromagnetic simulations to determine at which applied external magnetic field a given nanomagnet is flipped depending on the magnetic configuration of one of its neighbors. 
The field required to initiate reversal is then determined for an antiferromagnetic and for a ferromagnetic alignment of the two nanomagnets. 
To speed up convergence of the simulations, the damping parameter has been set to one; the values deduced from the simulations then overestimate the real reversal fields. 
However, the purpose of these simulations is not to quantify the reversal field but rather to determine which of the two configurations is more stable under an applied external magnetic field.

Results are reported in Fig.~\ref{fig:micromag} for two different cases: $s/L$=100\% [Fig.~\ref{fig:micromag}(a,b)] and $s/L$=80\% [Fig.~\ref{fig:micromag}(c,d)]. 
Although the two antiferromagnetic configurations [Fig.~\ref{fig:micromag}(a,c)] and the two ferromagnetic configurations [Fig.~\ref{fig:micromag}(b,d)] show strong similarities, the relative position of the nanomagnets' extremities has important consequences. 
When $s/L$=100\%, the two nanomagnets aligned ferromagnetically [Fig.~\ref{fig:micromag}(b)] are coupled via a magnetostatic field (black arrow) that goes against the external applied field.
Consequently, the total field felt locally by the two nanomagnets is weaker than the applied field and additional energy is necessary compared to the antiferromagnetic configuration [Fig.~\ref{fig:micromag}(a)] in order to initiate magnetization reversal.
The effect of this magnetostatic field has the opposite contribution when $s/L$=80\% and increases the total magnetic field felt locally by the nanomagnets [Fig.~\ref{fig:micromag}(d)].
In that case, the magnetostatic field favors magnetization reversal that occurs at an applied external field smaller than the one required in the antiferromagnetic configuration [Fig.~\ref{fig:micromag}(c)].
In other words, although the antiferromagnetic dimer state is the ground state configuration in both cases ($s/L$=100\%, $s/L$=80\%), this ground state is destabilized during the demagnetization protocol due to the magnetostatic coupling between the local micromagnetic texture within two neighboring nanomagnets. 
When $s/L$=80\%, a ferromagnetic configuration between neighboring elements is destabilized and a conventional antiferromagnetic state is favored. 
When $s/L$=100\%, the situation is reversed and a ferromagnetic alignment of neighboring nanomagnets is favored.
This result highlight the role of micromagnetism and the limitation of the Ising pseudo-spin approximation, as already suggested in other works where the curling of the magnetization at the nanomagnets' extremities is supposed to impact spin-flip events \cite{Zeissler2013, Rougemaille2013}.

To conclude, by tuning the vertical shift between neighboring nanomagnets arranged on a unidimensional chain, we observed magnetic configurations resulting from competing interactions. 
Besides the antiferromagnetic state and the dimer phase expected from the ANNNI model, we also evidenced a transition to an unexpected antiferromagnetic phase followed by a ferromagnetic state when the shift $s$ becomes higher than 50\%, typically. 
We believe that these two metastable states originate from the micromagnetic nature of the nanomagnets and the coupling of this additional magnetic degree of freedom with the applied external field during demagnetization. 
They are not expected in similar artificial spin chains that could be thermally activated. 
Our results also show that the condition $|J_1| / |J_2|$ = 2 can be achieved experimentally, thus allowing investigation of disordered and highly degenerated spin configurations in a one-dimensional system. 
Indeed, in that particular case, competing interactions destroy long-range order and lead for large distances to exponentially decaying spin-spin correlations superimposed with a spatial modulation \cite{Selke1988}. It would be interesting to further explore such artificial spin chains to test to what extent the ANNNI model correctly described the physics we observed. 

\begin{acknowledgements}
This work was partially supported by the Agence Nationale de la Recherche through Project No. ANR12-BS04-009 ''Frustrated''. The authors also thank the scientific staff of the Nanofab platform for the sample fabrication.

\end{acknowledgements}

\end{document}